\documentclass[sigconf]{acmart}
\AtBeginDocument{%
  }

\usepackage{algorithm}
\usepackage{algpseudocode}
\usepackage{graphicx}
\usepackage{textcomp}
\usepackage{xcolor}
\usepackage{colortbl}
\usepackage{xspace}
\xspaceaddexceptions{]\}}
\usepackage{listings}
\usepackage[switch]{lineno}
\usepackage{caption}
\usepackage[para]{footmisc}
\usepackage{hyperref}
\usepackage[inline,shortlabels]{enumitem}
\usepackage{subcaption}
\usepackage{wrapfig}
\usepackage{placeins}
\usepackage{epigraph}
\usepackage{fontawesome5}

\setcopyright{none}
\newcommand{\tname}[1]{\textsc{#1}\xspace}
\newcommand{\sysraw}{Aporia}
\newcommand{\sysinquote}{\ifhighlighttools{\setlength{\fboxsep}{1pt}\colorbox{aporia-highlight}{\textsc{Aporia}}}\else\textsc{Aporia}\fi}
\newcommand{\sysplain}{\textsc{\sysraw}\xspace}
\newcommand{\sys}{\ifhighlighttools{\setlength{\fboxsep}{1pt}\colorbox{aporia-highlight}{\textsc{\sysraw}}}\else\textsc{\sysraw}\fi\xspace}
\newcommand{\claudeplain}{\textsc{Claude Code}\xspace}
\newcommand{\claude}{\ifhighlighttools{\setlength{\fboxsep}{1pt}\colorbox{claude-highlight}{\textsc{Claude Code}}}\else\textsc{Claude Code}\fi\xspace}
\newcommand{\claudeinquote}{\ifhighlighttools{\setlength{\fboxsep}{1pt}\colorbox{claude-highlight}{\textsc{Claude Code}}}\else\textsc{Claude Code}\fi}
\newcommand{\notcrp}{\tname{NotCRP}}
\newcommand{\username}{Nikos\xspace}

\graphicspath{{figures/}}

\definecolor{aporia-purple}{HTML}{A781FE}
\definecolor{claude-orange}{HTML}{E8956A}
\definecolor{assistant-purple}{HTML}{6849B0}
\definecolor{human-gray}{HTML}{5D5D5D}
\colorlet{aporia-highlight}{aporia-purple!20}
\colorlet{claude-highlight}{claude-orange!30}
\newif\ifhighlighttools
\highlighttoolsfalse

\usepackage{tikz}
\usepackage{pgfplots}
\pgfplotsset{compat=1.18}
\newcommand{\circledletter}[2][aporia-purple]{%
	\raisebox{0.2ex}{%
		\tikz[baseline=(char.base)]{
			\node[
			circle,
			fill=#1,
			text=white,
			font=\sffamily\bfseries\small,
			inner sep=0.3pt,
			minimum size=1em
			] (char) {#2};
		}%
	}%
}

\newcommand{\usercl}[2][human-gray]{%
	\raisebox{0.2ex}{%
		\tikz[baseline=(char.base)]{
			\node[
			circle,
			fill=#1,
			text=white,
			font=\sffamily\bfseries\small,
			inner sep=0.3pt,
			minimum size=1em
			] (char) {#2};
		}%
	}%
}

\newcommand{\assistantcl}[2][assistant-purple]{%
	\raisebox{0.2ex}{%
		\tikz[baseline=(char.base)]{
			\node[
			circle,
			fill=#1,
			text=white,
			font=\sffamily\bfseries\small,
			inner sep=0.3pt,
			minimum size=1em
			] (char) {#2};
		}%
	}%
}

\newcommand{\scode}[1]{{\texttt{\small #1}}}
\newcommand{\code}[1]{\texttt{#1}}

\definecolor{skcolor}{rgb}{0.1,0.7,0.8}   
\definecolor{rrcolor}{rgb}{1.0,0.0,0.0}   
\definecolor{npcolor}{rgb}{1.0,0.65,0.0}  
\definecolor{hgcolor}{rgb}{0.2,0.8,0.2}   
\definecolor{hrcolor}{rgb}{0.2,0.4,0.8}   
\definecolor{bncolor}{rgb}{0.6,0.2,0.8}   
\definecolor{srcolor}{rgb}{0.9,0.1,0.6}   

\newcommand{\eqcontribmark}{%
  \textsuperscript{\normalfont\fontsize{10}{12}\selectfont\raisebox{-0.6ex}{*}}}

\newif\ifhidecomments
\hidecommentstrue

\newcommand{\eg}{\textit{e.g.,}\xspace}

\newcommand{\ie}{\textit{i.e.,}\xspace}
\newcommand{\vs}{\textit{vs.\@}\xspace}

\newcommand{\q}[1]{\emph{``#1''}}

\begin{document}

\title{Decision-Oriented Programming with \sysplain}
\thanks{* Equal contribution.}

\author{Saketh Ram Kasibatla\eqcontribmark}
\affiliation{%
  \institution{UC San Diego}
  \city{La Jolla}
  \state{CA}
  \country{USA}}
\email{skasibatla@ucsd.edu}

\author{Raven Rothkopf\,\eqcontribmark}
\affiliation{%
  \institution{UC San Diego}
  \city{La Jolla}
  \state{CA}
  \country{USA}}
\email{rrothkopf@ucsd.edu}

\author{Hila Peleg}
\affiliation{%
  \institution{Technion}
  \city{Haifa}
  \country{Israel}}
\email{hilap@cs.technion.ac.il}

\author{Benjamin C. Pierce}
\affiliation{%
  \institution{University of Pennsylvania}
  \city{Philadelphia}
  \state{PA}
  \country{USA}}
\email{bcpierce@cis.upenn.edu}

\author{Sorin Lerner}
\affiliation{%
  \institution{Cornell University}
  \city{Ithaca}
  \state{NY}
  \country{USA}}
\email{sorin.lerner@cornell.edu}

\author{Harrison Goldstein}
\affiliation{%
  \institution{University at Buffalo, SUNY}
  \city{Buffalo}
  \state{NY}
  \country{USA}}
\email{hgoldste@buffalo.edu}

\author{Nadia Polikarpova}
\affiliation{%
  \institution{UC San Diego}
  \city{La Jolla}
  \state{CA}
  \country{USA}}
\email{npolikarpova@ucsd.edu}

\renewcommand{\shortauthors}{Kasibatla et al.}

\begin{abstract}

AI agents help developers express computational intent abstractly, but at a cost:
developers cede decision-making authority to agents, often without realizing
that important design decisions are being made without them.
We bring these decisions to the foreground in a workflow we dub
\emph{decision-oriented programming}.
In DOP, decisions are (1) \emph{explicit and structured},
serving as the shared medium between programmer and agent;
(2) \emph{co-authored interactively},
with the agent eliciting them from the programmer;
and (3) \emph{traceable to code}.
As a step towards this vision, we built \sys, a
design probe that
tracks decisions in a persistent, editable Decision Bank;
elicits them proactively through design questions;
and encodes each as a test suite
that can help validate the agent's implementation.
A study of $14$ programmers showed that structuring collaboration around decisions kept programmer and agent in sync throughout the design process,
and highlighted opportunities to better support programmer decision-making in future systems.

\end{abstract}



\keywords{}

 \begin{teaserfigure}
 	\centering
   \includegraphics[width=0.95\textwidth]{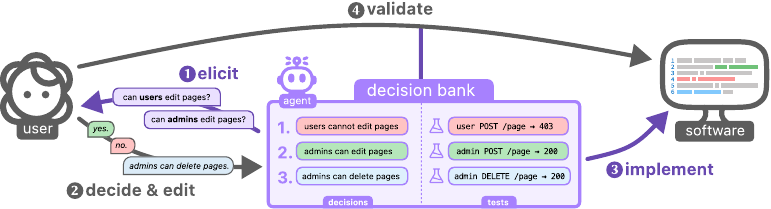}
   \caption{
      We describe Decision-Oriented Programming, a workflow that supports human decision-making in AI-assisted programming, and its instantiation in a system called \sys.
      In \sys, the programmer communicates with the agent via a structured representation of design decisions, called a \emph{Decision Bank}.
      The agent elicits decisions from the programmer~\assistantcl{1}, and the programmer responds and manages the decisions~\usercl{2}.
      The agent then uses the Decision Bank to generate code~\assistantcl{3}, and the programmer relies on the bank when validating the system behavior~\usercl{4}.
    }
   \label{fig:teaser}
 \end{teaserfigure}

\maketitle
\section{Introduction}
\label{sec:intro}
AI is changing the way people think about programming. This shift
began with GitHub Copilot~\cite{copilot} in 2022, and as of 2025, 84\% of developers
used AI tools, 47\% daily~\cite{stack-overflow-developer-survey}.
Modern tools go beyond auto-completion, with \emph{AI
agents}~\cite{claude-code, cursor, windsurf} autonomously
modifying codebases, executing shell commands, and implementing features end-to-end.

Studies exploring developer use of these tools report
	that using natural language prompts and letting AI deal with implementation details reduces cognitive effort~\cite{good-vibrations}
	and that using agents can help developers achieve flow states~\cite{good-vibrations,fawzy-vibe-coding}
	and improve productivity~\cite{good-vibrations,fawzy-vibe-coding,huang-2025}.
These experiences demonstrate the potential of AI programming tools to further a
fundamental design goal of programming languages: \emph{to express computation
at a level of abstraction that is closer to the way humans think and talk}~\cite{hopper-education, hopper-standardization}.

The downside of expressing intent abstractly is that crucial details are left out. 
Without explicit intervention, agents silently fill in these details, and researchers have observed that ``vibe coding'' often means ceding decision-making authority to agents altogether~\cite{good-vibrations}.
This problem is especially pronounced in open-ended tasks like software design, where requirements, and thus programmers' prompts, are underspecified.
Practitioners are increasingly worried about
the resulting \emph{cognitive debt}~\cite{echoes,cognitiveintentdebt}:
the erosion of a team's understanding and mental models of their own system.

Some professional developers have adopted workarounds by meticulously authoring implementation plans that specify major design decisions upfront before proceeding with code generation~\cite{huang-2025}.
Others rely on tools like Claude Code's Plan Mode~\cite{claude_plan_mode} and the specification-driven assistant Kiro~\cite{kiro}, where the agent generates and maintains design documents for developers to review alongside the code.
However, studies show that programmers struggle to check the agent's code against these plans~\cite{he2025plan}, and lose track of the decisions they have communicated to the agent in walls of prose, especially after multiple revisions~\cite{stepwise}.

To promote human decision-making during agentic software design, we propose to reify programmer decisions in a workflow we dub \emph{decision-oriented programming} (DOP,~\autoref{fig:teaser},~\autoref{sec:dop}).
In DOP, decisions are first-class objects from which software design is composed, and are automatically tracked and traceable to code.

As a foray into decision-oriented programming, we have built a
design probe called \sys\footnote{From the socratic term referring to the
  suspension of views on a subject, accompanied by a continuous questioning in
  an effort to better understand.}, which implements DOP.
It \emph{tracks} discrete decisions about system behavior in a persistent, structured plan (or Decision Bank) that programmers can view and edit at any time via a UI;
\emph{elicits} decisions by asking programmers \emph{design questions} about the software they are seeking to build; and
encodes each decision formally in a \emph{test suite},
giving users an unambiguous interpretation of their decisions
and providing executable feedback that helps validate the agent's implementation (\autoref{fig:user-scenario}).
Critically, \sys imposes a consistent, parallel structure across a decision's three forms---the design question that surfaces it, its entry in the Decision Bank, and the test suite that validates it---encouraging users to reason in the unified language of decisions across every stage of program design.

To understand the benefits and limitations of DOP, we conducted a within-subjects user study in which $N=14$ participants designed and implemented new features in a codebase, a setting representative of how developers incrementally extend real systems. Participants performed this task with the help of either \sys or a baseline planning agent.
We found that \sys kept participant intent and agent implementation in sync by scaffolding their interactions around discrete decisions, which produced a consistent structure throughout their sessions.
Participants used this structure to track their decisions, revisit them as designs evolved, and check the agent's implementation against them.
As a result, participants' mental models matched the implemented code more closely than when working with the baseline agent.
These findings suggest that structuring program design around explicit decisions is a promising way to keep programmers engaged with and in control of their designs.

Our study also surfaced several opportunities to better support human decision-making when designing software with agents.
We observed that decision elicitation is most effective when the programmers are presented with questions that align with their thinking and are grounded in the codebase,
and while front-loading decision-making helps programmers think more intentionally about their plans, it must be balanced with mechanisms for reconciling and iterating on decisions deeper in development.

The contributions of this paper are:
\begin{enumerate}
    \item articulating \emph{decision-oriented programming}, a workflow 
	which supports programmer decision-making in AI-assisted software design
    \item instantiating DOP with \sys,
	a design probe that elicits decisions through questions
	and formalizes those decisions in test suites,
	giving programmers a concrete way to validate that their intent is reflected in the final product
    \item evaluating \sys through a user study, demonstrating that DOP promotes programmer engagement in the design process
	and helps scaffold both design and validation
\end{enumerate}

\section{Related Work}\label{sec:related-work}

\paragraph{Empirical studies of AI assistants}

Recent research has examined users' interactions with AI agents~\cite{pu2025assistance}.
Pimenova et. al.~\cite{good-vibrations} note that at their best, programming agents can support flow, but that agentic programming has many pain points, such as difficulty communicating intent.
Fawzy et. al.~\cite{fawzy-vibe-coding} find that coding with agents has a speed-quality tradeoff, where developers use agents for enhanced flow, but overlook common quality assurance practices.
Sarkar and Drosos~\cite{sarkar-vibe-coding} identify an iterative process between prompting, reviewing, and manually editing AI-generated code. 
They note a shift of programmer expertise to evaluating outputs and deciding when to transition between prompting and editing, rather than writing code from scratch.
Huang et. al.~\cite{huang-2025} study how experienced developers use AI agents. They observe that developers adopt a variety of strategies to exercise agency while programming with AI, and suggest that carefully designed interfaces can help guide software developers to interact with agents more productively.
We believe that DOP can serve such a role, keeping agents' benefits while mitigating the pitfalls described in these studies.

\paragraph{Eliciting and exploring design intent}

A growing body of work tackles the challenge of helping users
clarify and communicate their intent to AI systems.
Several systems support \emph{intention formation} by involving users in an interactive process of refining goals.
\tname{CoLadder}~\cite{coladder} supports users generating code with LLMs by providing a UI for structuring prompts hierarchically.
%
\tname{Stepwise}~\cite{stepwise} facilitates data science practitioners using LLMs by providing opportunities for intervention and having AI state its assumptions at each step.
%
\tname{Cocoa}~\cite{cocoa} supports the development of research ideas. The tool centers around a plan document, which humans and AI collaboratively draft and execute.

Other works focus on \emph{design space exploration}.
For natural language generation, PDC~\cite{pdc} and \tname{Luminate}~\cite{luminate} help users organize and compare a range of different outputs from the same prompt.
%
\tname{HilDE}~\cite{hilde} uses a code-completion style UI to highlight critical decisions to users, allowing them to explore alternatives and discover their implicit intent.
Domain-specific tools such as \tname{Biscuit}~\cite{biscuit} and \tname{DynaVis}~\cite{dynavis}
provide a lightweight method to explore alternatives (\eg multiple options with a color picker) by dynamically generating UI widgets.
In \sys, instead of visualizing alternatives, users explore the design space through questions.

The system closest to ours, \tname{Pail}~\cite{pail}, pioneered an interaction mode for supporting AI-assisted design activities at higher levels of abstraction. 
\tname{Pail} asks users design questions, extracts and tracks requirements, and identifies implicit decisions made by the LLM.
We build on \tname{Pail}'s insight that tracking implicit and explicit decisions is valuable, and explore that space in-depth by making decisions the organizing unit of the co-design process.
\tname{Pail} fragments programmer intent across four separate views apart from the code and conversation, and its users struggled with information overload during studies~\cite{pail}.
\sys instead centers the workflow on a single concept, the decision, letting programmers work through a consistent structure as they discover, reify, and validate their intent against the code.


\paragraph{Structured intent representation}

Eliciting intent is complementary to \emph{representing} intent in a form that is convenient to track and revise.
\tname{Semantic Commit}~\cite{semantic-commit} presents a structured representation of agents' ``memories'' as sets of atomic specifications, that users can reconcile when in conflict.
We were inspired by this system when designing \sys's Decision Bank.
\tname{NeuroSync}~\cite{neurosync} externalizes the LLM's task decomposition as a manipulable graph, giving users a visual representation of intent.
\tname{CoLadder} and \tname{Stepwise}, the tools for hierarchical goal decomposition mentioned above~\cite{coladder, stepwise}
also feature a structured user interface that the user can inspect and revise.
\sys centers programming with AI agents around the Decision Bank, a structured representation of intent in the context of a codebase.


\paragraph{Testing strategies}
In Test-driven development (TDD)~\cite{beck2002driven, tdd-by-example}, expected behavior of code is formalized in the form of tests before any code is written.
Behavior-driven development (BDD)~\cite{bdd-review} is an evolution of TDD that focuses on higher-level concerns like overall system behavior. 
Property-based testing (PBT)  is a technique for testing executable properties of code with random inputs~\cite{quickcheck}. Inspired by PBT's emphasis on properties, \sys's tests are organized into suites grouped by the decisions they validate rather than by components under test.
These testing strategies have been explored as a means to improve correctness of AI programming agents~\cite{tdflow, tenet, classeval, bose2025prompts, vikram2023can}, but not with developers in the loop.
\tname{TiCoder}~\cite{ticoder} formalizes user intent with LLMs by generating test cases and asking users to confirm or deny whether each matches their intention.
\sys also generates test cases with a human in the loop. 
However, \sys does so using AI agents, which operate on the scale of a whole codebase rather than individual functions.
\section{Design Motivations}\label{sec:dop}
We first introduce decision-oriented programming (DOP, \autoref{fig:teaser}) for agentic software design, which is characterized by three design goals motivated by a review of HCI and Interaction Design literature:
\begin{enumerate}\setlength{\itemsep}{\baselineskip}
\item[\textbf{DG1}] \textbf{Decisions are explicit and structured.} \label{dg:DG1}
Our approach is inspired by the tradition of design rationale, which argues for structured representations
of the reasoning behind design choices~\cite{ibis,moran1996design,qoc}, and makes design reasoning explicit and persistent as a ``coproduct of design''.
Specifically, Questions, Options, Criteria (QOC) notation~\cite{qoc} represents design spaces in terms of \emph{Questions} that identify key issues,
\emph{Options} that provide possible answers, and \emph{Criteria} for evaluating those options. 
In DOP, rather than being buried in multiple revisions of minimally structured plans, programmer decisions are reified as first-class objects in a persistent, editable record;
this record is the shared medium through which programmer and agent negotiate the design.
\item[\textbf{DG2}] \textbf{Decisions are co-authored interactively.} \label{dg:DG2}
Interfaces with deliberate micro-level frictions have been shown to disrupt mindless automatic interactions and prompt moments of reflection~\cite{design_frictions}, and these reflections are integral to the design process~\cite{buxton2010sketching}.
Micro-level frictions are also a useful mechanism for promoting critical thinking in AI-assisted tools~\cite{kazemitabaar2025exploring, gould2024chattl}.
In DOP, the agent proactively elicits decisions from the programmer, introducing friction that keeps them engaged rather than passively reviewing agent-generated plans.
\item[\textbf{DG3}] \textbf{Decisions are traceable to code.} \label{dg:DG3}
Users struggle to form concrete intentions when they use generative AI in place of other design processes~\cite{gulf_of_envisioning}.
One factor of in this phenomenon, otherwise known as the ``gulf of envisioning'' is that users lack internal ``cognitive benchmarks,'' or concrete expectations for what a correct output should look like, to anchor their validation. 
Surveys report that professional developers using agents often resort to creating these benchmarks themselves, but doing so requires discipline and workarounds like restricting the scope of each agent change, reviewing generated code line by line, and manually authoring plans and design documents~\cite{huang-2025,sarkar-vibe-coding}.
In DOP, each user decision is tied to a specification of how the code should behave to satisfy that decision, supplying programmers with benchmarks to help validate that agent-generated code reflects their intent.
\end{enumerate}

\autoref{sec:design-considerations} describes how we implemented these design goals in the \sys programming assistant.
\section{The \sys programming assistant}\label{sec:aporia}

Our goal is to support human decision-making in the context of
agentic planning.
To this end, we implemented the \sys programming assistant
to explore the DOP workflow.
This section demonstrates \sys by example
and describes the design considerations behind the tool and its implementation.
%






\begin{figure*}[t]
    \centering
    \includegraphics[width=0.95\textwidth]{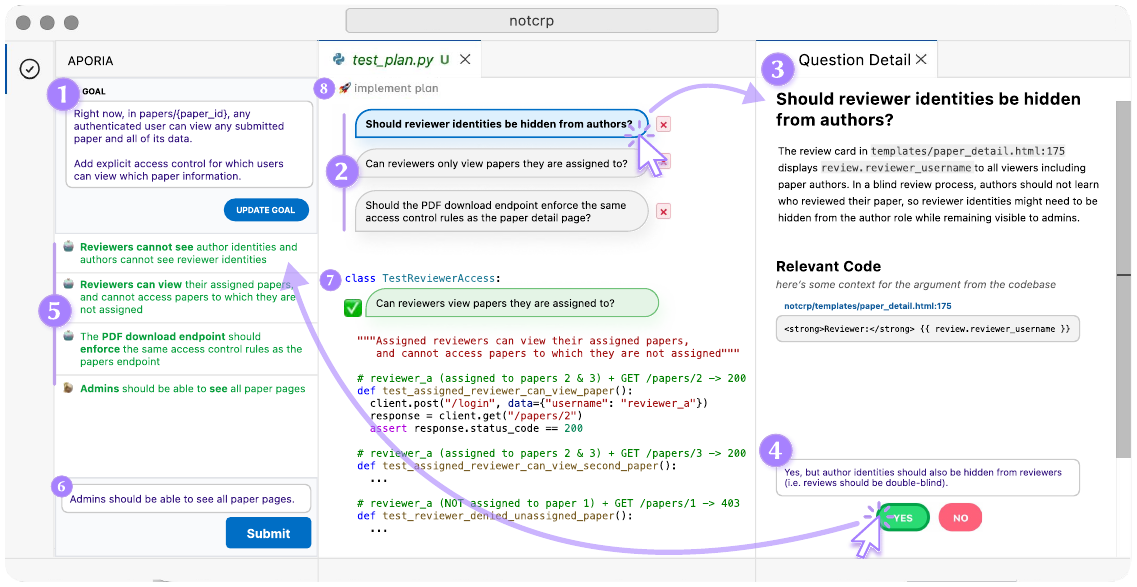}
    \captionof{figure}{
\sys is a VS Code extension that supports decision-oriented programming.
Users enter their high-level goal into the Goal Field~\circledletter{1}, which prompts \sys to generate design questions as Question Bubbles~\circledletter{2}.
Clicking on a question opens the Question Detail View~\circledletter{3}, which includes \sys's argument for the question, as well as references to relevant code.
Users can answer the question with a yes/no response and an optional comment~\circledletter{4}.
Decisions are accumulated in the Decision Bank~\circledletter{5}, where they can be edited or revoked.
Users can also add decisions that were not elicited by \sys via the Custom Decision field~\circledletter{6}.
Each decision is formally encoded in a test suite that validates the implementation against that decision~\circledletter{7}.
Once users are satisfied with their design, they can click ``Implement plan''~\circledletter{8} to prompt \sys to implement the feature in the codebase.
    }
    \label{fig:user-scenario}
\end{figure*}

    \subsection{User Scenario}\label{sec:user-scenario}

\begin{figure}[t]
  \centering
  \includegraphics[width=\columnwidth]{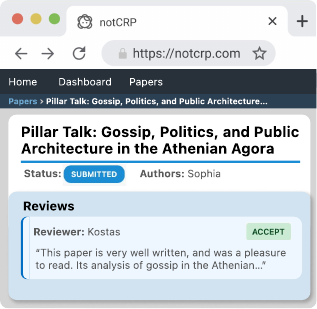}
  \caption{An example paper details page in the \notcrp conference management web app.}
  \label{fig:notcrp-paper-details}
  \Description{A paper details page in the NotCRP conference management web app.}
\end{figure}

Frustrated with the poor usability of existing conference management systems,
\username decided to build his own, dubbed \notcrp, for a conference he's running.
\username has a basic Python web app already in place:
\autoref{fig:notcrp-paper-details} shows an example page from the app
for displaying all the details of a chosen paper.
He now wants to add {access control} to this page,
restricting which users can see which details.
\username starts by opening the \sys sidebar in his editor,
and typing in his high-level goal into the \emph{Goal Field}
at the top of the sidebar (\circledletter{1} in \autoref{fig:user-scenario}). 

\paragraph{Eliciting decisions via question generation}
%
In response, \sys generates yes/no questions meant to elicit design decisions relevant to the goal
(in this case, about the access control policy);
these appear as \emph{Question Bubbles}~(\circledletter{2}) in the editor, inside an \sys-managed plan file.
\username can decide to engage with or dismiss any question.
In this example, he clicks on the first question (``Should reviewer identities be hidden from authors?''),
which opens the \emph{Question Detail View}~(\circledletter{3}) on the right,
where he can see \sys's rationale for answering ``yes'' \vs ``no'' underneath the Detail View header.
(``In a blind review process, authors should not learn who reviewed their paper...'').
The Detail View also contains clickable references to relevant code.

\username likes the idea of making the review process blind;
moreover, the question reminds him that in fact he wanted to enforce a \emph{double-}blind review process,
where reviewers also don't learn author identities.
He enters this as a comment into the response field~(\circledletter{4})
and clicks ``yes'' to answer the question positively.

\paragraph{Persistent decisions via the Decision Bank}
\username's decision to make the review process double-blind now appears in \sys's \emph{Decision Bank}~(\circledletter{5}) below the goal.
This allows \username to keep track of the decisions he's made
and easily change his mind later by clicking on the decision and editing his answer.
He answers a few more questions, and his decisions accumulate in the Bank.
At some point, it occurs to him that admins should be always able to see all the details of all papers;
this is not something \sys had asked about,
but he can add this decision to the Bank via the Custom Decision field~(\circledletter{6}).

\paragraph{Traceability via test suites}
Every time \username commits a decision to the Bank,
\sys generates a test suite to validate the (future) implementation against that decision.
\autoref{fig:user-scenario}~(\circledletter{7}) shows the test suite generated for the double-blind review decision.
\username sees that each test case is preceded by a natural-language comment summarizing its inputs and outputs and can inspect any test case if he wants to see what exactly is being checked.

\paragraph{Implementation and validation}
Re-reading the decisions in the Bank, \username feels that the design is now sufficiently well-defined,
so he clicks the ``Implement plan'' code lens at the top of the plan file~(\circledletter{8}),
which signals \sys to modify the \notcrp codebase to implement the goal.
During this process, \sys automatically runs all its generated tests
to validate the implementation against the user's decisions.
When it finishes, \username runs the app and evaluates its behavior manually,
using the Decision Bank as a checklist
of his desired access control policies.


\paragraph{Comparison with traditional planning agents}

In contrast to \sys, state-of-the-art coding agents~\cite{claude-code, cursor,
  windsurf} typically organize their design iteration with users around
natural-language \emph{plans}.%
\footnote{Although some commercial agents also incorporate questions into their
  interaction model, questions do not play a central role; we describe a pure
  plan-oriented workflow here for the sake of comparison.}
Imagine that \username uses a traditional planning agent instead of \sys for the same task.
He prompts the agent with the same high-level goal
and in response receives a plan:
a linear document of natural language and pseudocode
describing the design decisions the agent made---for
example, ``Reviewer identities should be hidden from authors to ensure a blind review process.''
\username reviews the plan, but the design decisions are buried in a long document,
and the easiest action is to accept the plan as-is.
He thus doesn't notice that the agent chose a single-blind review process---or even
that single- \vs double-blind is something worth thinking about.

\subsection{Design}
\label{sec:design-considerations}

As demonstrated in the previous section, \sys operationalizes the design goals of DOP (\autoref{sec:dop}) through three main affordances:
the Decision Bank accumulates decisions as editable entries, making them \emph{explicit and structured} (\hyperref[dg:DG1]{\textbf{DG1}}); 
the Question Bubbles surface design questions that actively engage the user in the design process and \emph{elicit decisions} (\hyperref[dg:DG2]{\textbf{DG2}});
and the per-decision test suites make decisions \emph{traceable to code} (\hyperref[dg:DG3]{\textbf{DG3}}), lowering the effort involved in validating the implementation.
In this section, we first define what constitutes a design decision in \sys, then share some key lessons we learned through pilot testing that sharpened these affordances.

\subsubsection{Decisions in \sys}
\label{sec:decision-def}
In \sys, decisions are represented as atomic, generally orthogonal choices about the behavior of a feature the user intends to implement, for example, whether a paper's reviewer identities should be hidden from its authors (\autoref{sec:user-scenario}). 
The scope of a decision must be small enough for a user to understand its implications at a glance and concrete enough to encode as a check on the code.
Together, the decisions accumulated under a high-level goal for a feature specify how the feature should be implemented in the context of the codebase.

As mentioned in~\autoref{sec:dop}, \sys elicits these decisions via QOC notation~\cite{qoc},
which structures design rationale around
\emph{Questions} that identify key issues (our design questions presented in the Question Bubbles),
\emph{Options} that provide possible answers (yes and no answers, in our case), and
\emph{Criteria} for evaluating those options (our arguments and relevant code in the Question Detail View).

\subsubsection{Additional design lessons}
Through several pilot studies, we iterated on the basic design of our Decision Bank, Question Bubbles, and tests and arrived at four key lessons.

\paragraph{Parallel structure between questions, decisions, and test suites:}
To address observed difficulties connecting questions to their corresponding decisions, we introduced a consistent parallel structure across three levels of our interface.
First, we rewrote decision titles to match the linguistic structure of the questions that produced them.
Second, inspired by property-based testing~\cite{quickcheck}, we reorganized
tests into suites grouped by the {decisions} they validate rather than the
components they test.
Finally, we preceded each test suite with its corresponding Question Bubble, making it easier to trace the connections between the two.

\paragraph{Progressive disclosure:} 
We found that displaying too many questions at once overwhelmed our pilot
participants, while too few left them with idle time and broke their flow, so we settled on revealing
five at a time.
We reached a similar balance of friction with the generated test suites.
Showing each test in full overwhelmed pilot participants when trying to understand them,
so we added a short summary of each test case's inputs and outputs (\eg \code{reviewer\_a
  (assigned to papers 2 \& 3) + GET /papers/3 -> 200})
in a preceding comment. These summaries show the 
gist of what is being tested while test bodies are collapsed in the editor.

\paragraph{Binary questions reduce overload:}
Early prototypes required participants to write their own responses to all design questions; pilots showed that yes/no questions with optional user comments struck the right balance between expressiveness and cognitive cost.

\paragraph{Grounding in code:}
Programmers wanted to consider questions in the context of the codebase rather
than in the abstract, which prompted us to include clickable relevant code references in
the Detail View.

\subsection{Implementation}
\label{sec:implementation}
\sys is implemented as an extension for Visual Studio Code~\cite{visual-studio-code} and is written in Typescript. All UI elements are implemented using React. The \sys sidebar and Question Detail View are displayed in \code{Webview}s, while Question Bubbles are displayed in \emph{\code{WebviewEditorInset}s} (only available in the insiders' build).

\sys orchestrates three specialized agents---a \emph{questioner}, a \emph{planner}, and an \emph{implementer}---each an instance of \claude controlled using the Agent Client Protocol~\cite{acp}.
The questioner generates design questions and supporting arguments based on the goal; the planner formalizes decisions into test suites; and the implementer changes the codebase in line with the goal, decisions, and tests. Agents run concurrently, with \sys managing asynchronous state updates by queueing and batching requests that are made while an agent is working.

Agents communicate with the extension via a custom Model Context Protocol (MCP)~\cite{mcp} server featuring two tools. 
\code{submit\_argument} stores questions and arguments in a shared database, which the extension renders as Question Bubbles.
\code{submit\_uuid\_\allowbreak to\_test\_\allowbreak suite\_mapping}, is used by the planner to map test suite names (\eg \code{TestReviewerAccess} in \autoref{fig:user-scenario}) to decisions. The extension uses this mapping to render Question Bubbles for each decision next to its test suite (\eg the green Question Bubble in~\autoref{fig:user-scenario}).

\section{User Study}
\label{sec:study}
We describe the design of our user study, which was intended to answer two
research questions:


\begin{enumerate}
    \item[\textbf{RQ1}] How does \sys affect developers' ability to \emph{discover and articulate} the design decisions relevant to a given programming task?
    \item[\textbf{RQ2}] How does \sys affect developers' \emph{perceived and actual
      understanding} of the agent's implementation and their strategies for
    validating their understanding?
\end{enumerate}

\subsection{Participants}
\label{sec:participants}
%
We recruited 14 participants, 8 self-identified as men and 6 as women.
2 were undergraduate students, 6 were graduate students, and 6 were professional software engineers.
We required participants to have at least some familiarity working with Python and AI agents.
All reported moderate to high Python proficiency and said they used AI
programming agents at least a few times a week.

\definecolor{cellgreen}{HTML}{C6EFCE}
\definecolor{cellred}{HTML}{FFC7CE}
\definecolor{cellgray}{HTML}{D9D9D9}
\definecolor{cellyellow}{HTML}{FFEB9C}

\begin{figure*}[t]
\footnotesize
\setlength{\tabcolsep}{3pt}
\renewcommand{\arraystretch}{1.2}
\arrayrulecolor{gray}

\begin{minipage}[t]{0.48\textwidth}
{\normalsize\textbf{Task A (Paper detail access control)}}

\vspace{4pt}

\faClipboardList\;\textbf{Survey Question:} In your implementation, which users (admins, authors, reviewers, other users) can see what Paper data (pdf, status, paper authors, assigned reviewers, reviews) and when can they see it?

\vspace{6pt}

\faUser\;\textbf{P12's answer:}\par\vspace{2pt}
{\color{gray}\vrule width 2pt}\hspace{6pt}\begin{minipage}[t]{0.92\linewidth}\emph{``Admins can see everything. Authors can see all paper details, except pending reviews. Reviewers can't see author names, but they can see other reviewers' submissions. Unrelated users can see nothing.''}\end{minipage}

\vspace{6pt}
{\color{gray}\faArrowDown\;\textbf{Encoded as:}}
\vspace{4pt}

\begin{minipage}[t]{0.48\linewidth}
\resizebox{\linewidth}{!}{%
\begin{tabular}{@{}l
  >{\centering\arraybackslash}p{2em}
  >{\centering\arraybackslash}p{2em}
  >{\centering\arraybackslash}p{2em}
  >{\centering\arraybackslash}p{2em}@{}}
\multicolumn{5}{@{}l}{\textbf{Description Grid}} \\
\toprule
& Adm. & Auth. & Rev. & Unrel. \\
\midrule
PDF            & Yes & Yes & Yes & No \\
Status         & Yes & Yes & N/D & No \\
Authors        & Yes & Yes & No  & No \\
Reviewers      & Yes & Yes & N/D & No \\
Reviews        & Yes & No  & Yes & No \\
\bottomrule
\end{tabular}
}
\end{minipage}%
\hfill
\begin{minipage}[t]{0.48\linewidth}
\resizebox{\linewidth}{!}{%
\begin{tabular}{@{}l
  >{\centering\arraybackslash}p{2em}
  >{\centering\arraybackslash}p{2em}
  >{\centering\arraybackslash}p{2em}
  >{\centering\arraybackslash}p{2em}@{}}
\multicolumn{5}{@{}l}{\textbf{Implementation Grid}} \\
\toprule
& Adm. & Auth. & Rev. & Unrel. \\
\midrule
PDF            & \cellcolor{cellgreen}Yes & \cellcolor{cellgreen}Yes & \cellcolor{cellgreen}Yes & \cellcolor{cellred}Yes \\
Status         & \cellcolor{cellgreen}Yes & \cellcolor{cellgreen}Yes & \cellcolor{cellgray}Yes & \cellcolor{cellgreen}No \\
Authors   & \cellcolor{cellgreen}Yes & \cellcolor{cellgreen}Yes & \cellcolor{cellgreen}No  & \cellcolor{cellgreen}No \\
Reviewers & \cellcolor{cellgreen}Yes & \cellcolor{cellgreen}Yes & \cellcolor{cellgray}Yes & \cellcolor{cellgreen}No \\
Reviews        & \cellcolor{cellgreen}Yes & \cellcolor{cellgreen}No  & \cellcolor{cellgreen}Yes & \cellcolor{cellgreen}No \\
\bottomrule
\end{tabular}
}
\end{minipage}
\end{minipage}%
\hfill
{\color{gray}\vrule}
\hfill
\begin{minipage}[t]{0.48\textwidth}
{\normalsize\textbf{Task B (Reviewer assignment algorithm)}}

\vspace{4pt}

\faClipboardList\;\textbf{Survey Question:} In your implementation, which users (admins, users) can be assigned to review which papers, and under what conditions (author/co-author/institutional conflicts, workload, already assigned, etc.) is assignment prevented?

\vspace{6pt}

\faUser\;\textbf{P3's answer:}\par\vspace{2pt}
{\color{gray}\vrule width 2pt}\hspace{6pt}\begin{minipage}[t]{0.92\linewidth}\emph{``Authors can't be assigned to the papers they submitted, but they can be assigned to papers from others; reviewers from the same university as the authors can't be assigned to their papers.''}\end{minipage}

\vspace{6pt}
{\color{gray}\faArrowDown\;\textbf{Encoded as:}}
\vspace{4pt}

\begin{minipage}[t]{0.38\linewidth}
\resizebox{\linewidth}{!}{%
\begin{tabular}{@{}l l@{}}
\multicolumn{2}{@{}l}{\textbf{Description Grid}} \\
\toprule
Eligibility & Ranking \\
\midrule
Not an author         & N/D \\
Different institution & \\
\addlinespace
\addlinespace
\addlinespace
\addlinespace
\addlinespace
\addlinespace
\bottomrule
\end{tabular}
}
\end{minipage}%
\hfill
\begin{minipage}[t]{0.58\linewidth}
\resizebox{\linewidth}{!}{%
\begin{tabular}{@{}l l@{}}
\multicolumn{2}{@{}l}{\textbf{Implementation Grid}} \\
\toprule
Eligibility & Ranking \\
\midrule
\cellcolor{cellgreen}Not an author                & \cellcolor{cellgray}Alphabetical \\
\cellcolor{cellgreen}Different institution         & \\
\cellcolor{cellred}Not author on other papers      & \\
\cellcolor{cellgray}Not already assigned           & \\
\addlinespace
\addlinespace
\bottomrule
\end{tabular}
}
\end{minipage}
\end{minipage}

\caption{Sample correctness analysis of post-task survey responses from P12 (Task A, left) and P3 (Task B, right).
Free-form answers are encoded into a Description Grid, which is then compared to the actual implementation encoded in the Implementation Grid.
Policy alignment is classified into {\setlength{\fboxsep}{1pt}\colorbox{cellgreen}{matches}}, {\setlength{\fboxsep}{1pt}\colorbox{cellred}{mismatches}}, or {\setlength{\fboxsep}{1pt}\colorbox{cellgray}{not described}} (N/D).
}
\label{fig:sample-correctness}
\end{figure*}

\subsection{Study Procedure}
\label{sec:procedure}

We conducted a comparative structured observation study~\cite{comparative-structured-observation},
where participants completed two programming tasks,
one with \sys and one with our baseline, \claude~\cite{AnthropicClaudeCode},
a state-of-the-art coding agent
which follows the plan-oriented workflow described in \autoref{sec:user-scenario}.
(Although it can ask clarifying questions, it does so sporadically.)

We ran a counterbalanced within-subjects study across two factors, assistant order (\sys first vs. \claude first) and task order (Task A first vs. Task B first), resulting in four configurations.
All studies were conducted remotely over Zoom and facilitated by one of the first two authors.
Participants completed tasks in a browser-based
\scode{code-server}~\cite{codeserver} VS Code IDE~\cite{vscode} running in an
isolated Docker container~\cite{merkel2014docker}.
Both assistants used the same underlying model, Claude Sonnet 4.6~\cite{claude_sonnet_4_6_2026}, to isolate the effect of \sys's interface.
Each session took approximately 90 minutes, and participants were compensated with a \$35  Amazon gift card.

Participants began their session by signing an informed consent form compliant with our IRB approval.
They were then given a brief introduction and told they would be completing programming tasks with two different AI programming assistants, renamed \scode{Assistant-1} (\sys) and \scode{Assistant-2} (\claude) to minimize bias.

\subsubsection{Assistant tutorial} 
Before each task, participants completed a 10-minute tutorial for their assigned assistant, illustrated with in-situ screenshots.
The tutorial remained as a reference for participants throughout the task.  

\subsubsection{Programming task and post-task survey}
Participants completed a 25-minute programming task with each 
assistant, modifying the \notcrp conference management system
(\autoref{sec:user-scenario}).
Each task was intentionally open ended, involving multiple subjective design
decisions:

\begin{description}
    \item [\textbf{Task A}]\emph{(Paper detail access control)}: ``Currently,
    any authenticated user can view every paper and all of its associated
    data. Modify the \scode{Paper} detail page to add explicit access control
    for which users can view which paper's information.''  (This task inspired
    our user scenario in~\autoref{sec:user-scenario}.)
    \item [\textbf{Task B}]\emph{(Reviewer assignment algorithm)}: ``Currently, the reviewer assignment form requires conference admins to manually type in a reviewer's username to assign them to a paper. Modify the form to display a filtered and ranked list of eligible reviewers for admins to select from.''
\end{description}

For each task, participants were given one of the task descriptions above, the \notcrp repository, a prepopulated database of users and papers for testing, and a running instance of \notcrp. After each task, participants completed a post-task survey that collected data as detailed in ~\autoref{sec:data_collection}.


\subsubsection{Post-study survey and semi-structured interview}
After finishing the tasks, participants completed a post-study
survey and a set of semi-structured interview questions asking them to reflect on their experience with each assistant.

\subsection{Data Collection and Analysis}
\label{sec:data_collection}

We now describe our methods for data collection and analysis, with our results to be found in~\autoref{sec:results}.

\subsubsection{Systematic mental review}
\label{sec:mental-review}
In designing our post-task surveys, we were interested in how participants' \emph{subjective confidence} in their solutions varied after they were guided through a systematic mental review of their work.
This review comprised a series of questions designed to probe aspects of their implementation that they might not have considered.
Participants were asked to list up to three \emph{design decisions} that shaped
their solution and, for each, say (a) what helped them realize there was a
decision to make and (b) who made the decision, from ``entirely myself'' (1) to
``entirely the assistant'' (5).
We also asked a \emph{policy question} (see the top part of \autoref{fig:sample-correctness}),
which prompted them to explain the access control policies they implemented for Task A and the reviewer eligibility policies they implemented for Task B;
we then asked which strategies they used to validate that the code indeed implemented those policies.

We assessed participants' confidence that their implementation matched their intent both before and after this review.
Critically, participants completed the surveys without access to their code or the app, so that responses reflected what they believed about their implementation, not what they were able to validate in the moment.
We also measured {cognitive load} using NASA-TLX~\cite{hart1988development}.

\subsubsection{Correctness analysis}
\label{sec:correctness}

We used the policy questions from the post-task survey to conduct a \emph{correctness analysis}
comparing participants' descriptions of their implemented policies to the actual behavior of their code.
\autoref{fig:sample-correctness} shows the full pipeline:
starting from the participant's natural-language description of the policies,
we manually encoded it into a Description Grid, where each cell represents an atomic policy.
For Task A, an atomic policy states whether a given role (Admin, Author, Reviewer, Unrelated user)
can view a given paper data (PDF, status, author names, reviewer names, reviews),
marked as ``Yes,'' ``No,'' or ``Conditional'' for policies with temporal constraints (\eg \emph{``authors can see reviews only after a decision has been made''}).
Task B's grid captures what properties make a reviewer eligible to review a paper and how reviewers are ranked.

We then independently encoded each participant's submitted code onto the same grid format (the Implementation grid) and compared the two.
Each grid cell was classified as either:
a {\setlength{\fboxsep}{1pt}\colorbox{cellgreen}{match}} if the two grids had equivalent values,
a {\setlength{\fboxsep}{1pt}\colorbox{cellred}{mismatch}} if they conflicted,
or {\setlength{\fboxsep}{1pt}\colorbox{cellgray}{not described}} if a cell in the Implementation grid had no corresponding cell in the Description grid
(\ie the behavior was present in the code but absent from the participant's
description
).

\subsubsection{Decision categorization}
\label{sec:decision-coding}

After conducting the studies, we used telemetry to identify moments where participants actively made decisions with each tool.
For \sys, we considered answering a question or entering a custom decision to be an active decision.
For \claude, we manually separated participants' messages and answers to questions into decisions of similar granularity to those made using \sys (as described in~\autoref{sec:design-considerations}), considering each to be a separate active decision.
We also distinguished \emph{elicited decisions} (\ie decisions made in response to a question from \sys or \claude) from decisions overall, which included participants' messages to \claude and custom decisions in \sys.

\subsubsection{Quantitative analysis}

To estimate the quantitative relation between the used assistant $a$ ($0$ for
\claude, $1$ for \sys) and each of the performance metrics while accounting for possible learning effects, we fit Generalized Linear Models (GLMs)~\cite{mcelreath2018statistical} using \code{pymer4}~\cite{jolly2018pymer4}.
Each GLM uses a performance metric $m$ as outcome variable, $a$ as main predictor, and the task (Task A or Task B) and whether it occurred first or second as further covariate predictors.
For performance metrics that are count data (\eg number of elicited decisions), the model uses a Poisson distribution with log link function;
for performance metrics that are ratios (\eg percent of described criteria that were successfully implemented), the model uses a Binomial distribution with log link function.

After fitting each GLM on the experimental data, we report the mean, p-value, and 95\% probability interval for the estimate of the coefficient of $a$ on the outcome scale (that is, after inverting the link function), which can be interpreted as an estimate of the effect of using \sys over \claude on the performance metric.
As is customary, we interpret a $p$-value below $0.05$ to denote a "significant" difference between the two sets of measures.

\subsubsection{Thematic analysis}
We recorded and transcribed each participant's session and interview.
They were encouraged to think aloud while completing each task, verbalizing their problem-solving process, reactions to questions, etc.
We used thematic analysis~\cite{braun2006using, vaismoradi2013content} to identify themes from the transcripts.
Two authors individually coded participant quotes from the transcripts related
to our research questions and collaboratively grouped these codes into broader
themes to present with our quantitative results.

\section{Results}
\label{sec:results}
\highlighttoolstrue
\looseness-1
Our results show how \sys helps developers make informed decisions and build a more accurate understanding of code. 
We present a qualitative and quantitative analysis answering two RQs (see~\autoref{sec:study}).

\definecolor{likert-sd}{HTML}{1E3A5F}   
\definecolor{likert-d}{HTML}{3D6B8E}    
\definecolor{likert-n}{HTML}{BFBFBF}    
\definecolor{likert-a}{HTML}{C9B2FE}    
\definecolor{likert-sa}{HTML}{A781FE}   

\begin{figure}[t]
\begin{flushleft}
\footnotesize
\faClipboardList\;\textbf{(Post-study) Survey Question:} I found the following feature of \sysplain useful:
\end{flushleft}

{\footnotesize
\tikz\fill[likert-sd] (0,0) rectangle (0.2cm,0.2cm);\,Strongly Disagree\quad
\tikz\fill[likert-d] (0,0) rectangle (0.2cm,0.2cm);\,Disagree\quad
\tikz\fill[likert-n] (0,0) rectangle (0.2cm,0.2cm);\,Neutral\quad
\tikz\fill[likert-a] (0,0) rectangle (0.2cm,0.2cm);\,Agree\quad
\tikz\fill[likert-sa] (0,0) rectangle (0.2cm,0.2cm);\,Strongly Agree
}

\vspace{4pt}

\newcommand{\likertbar}[5]{%
  \hspace{-0.5cm}\begin{tikzpicture}[baseline=0cm]
    \pgfmathsetmacro{\negoffset}{-#2*0.3-0.5*#3*0.3}
    \node[anchor=east, font=\footnotesize, text width=3cm, align=right] at (\negoffset cm-0.1cm, 0) {#1};
    \pgfmathsetmacro{\xpos}{-#2*0.3-0.5*#3*0.3}
    \ifnum#2>0
      \fill[likert-d] (\xpos cm, -0.12cm) rectangle ++(#2*0.3cm, 0.24cm);
      \node[font=\tiny, white] at (\xpos cm + #2*0.15cm, 0) {#2};
      \pgfmathsetmacro{\xpos}{\xpos+#2*0.3}
    \fi
    \ifnum#3>0
      \fill[likert-n] (\xpos cm, -0.12cm) rectangle ++(#3*0.3cm, 0.24cm);
      \node[font=\tiny] at (\xpos cm + #3*0.15cm, 0) {#3};
      \pgfmathsetmacro{\xpos}{\xpos+#3*0.3}
    \fi
    \ifnum#4>0
      \fill[likert-a] (\xpos cm, -0.12cm) rectangle ++(#4*0.3cm, 0.24cm);
      \node[font=\tiny, white] at (\xpos cm + #4*0.15cm, 0) {#4};
      \pgfmathsetmacro{\xpos}{\xpos+#4*0.3}
    \fi
    \ifnum#5>0
      \fill[likert-sa] (\xpos cm, -0.12cm) rectangle ++(#5*0.3cm, 0.24cm);
      \node[font=\tiny, white] at (\xpos cm + #5*0.15cm, 0) {#5};
      \pgfmathsetmacro{\xpos}{\xpos+#5*0.3}
    \fi
  \end{tikzpicture}%
}

\likertbar{Answering questions}{0}{4}{6}{5}
\par\vspace{2pt}
\likertbar{The generated test suites}{3}{7}{2}{3}
\par\vspace{2pt}
\likertbar{The Decision Bank}{0}{1}{7}{7}
\par

\caption{Distribution of participants' post-study survey likert responses after using \sysplain.}
\label{fig:likert-aporia}
\end{figure}

Overall, participants had positive general impressions of \sys's various affordances (see \autoref{fig:likert-aporia}).
The Decision Bank was rated most favorably ($M=4.4$/5), with all but one participant agreeing or strongly agreeing that it was useful.
The question bubbles were similarly well-received ($M=4.1$/5).
Participants were more divided on the test suites ($M=3.3$/5): three disagreed
and seven were neutral, a finding we explore further
in~\autoref{sec:DG3}.

Using Wilcoxon signed-rank tests~\cite{Wilcoxon1992} to compare NASA-TLX scores revealed no significant differences 
between \sys and \claude across any of the five subscales.
Participants reported similarly
low levels of mental demand ($M_{\sysplain}=2.50$, $M_{\tname{Claude}}=2.43$),
temporal demand ($M_{\sysplain}=2.64$, $M_{\tname{Claude}}=2.64$), effort
($M_{\sysplain}=1.86$, $M_{\tname{Claude}}=2.14$), and frustration ($M_{\sysplain}=1.79$, $M_{\tname{Claude}}=1.43$), and similarly high levels of perceived success ($M_{\sysplain}=3.71$, $M_{\tname{Claude}}=3.50$).
Task-level comparisons also
showed no significant
differences 
though Task B trended marginally toward higher effort ($M_{A}=1.79$, $M_{B}=2.21$; $p=0.058$).

\FloatBarrier

\subsection{RQ1: Discovering \& articulating decisions}
\label{sec:rq1}

\subsubsection{\sysplain helped participants discover design considerations through decision elicitation}
\label{sec:discover}

\begin{figure*}[t]
\begin{minipage}[t]{0.48\textwidth}
\vspace{0pt}
\begin{flushleft}
\footnotesize
\faClipboardList\;\textbf{Survey Question:} For each design decision you described, what helped you realize there was a decision to make there? (select up to two)

\vspace{2pt}
\end{flushleft}
\scriptsize
\tikz\fill[aporia-purple] (0,0) rectangle (0.2cm,0.2cm);\,\sysplain\quad \tikz\fill[claude-orange] (0,0) rectangle (0.2cm,0.2cm);\,\claudeplain\quad *User-written response

\vspace{4pt}
\hspace{-1cm}\begin{tikzpicture}
\begin{axis}[
    xbar stacked,
    width=0.7\columnwidth,
    height=5cm,
    bar width=8pt,
    xmin=0,
    xmax=42,
    ytick=data,
    yticklabels={
        {Prior knowledge or experience},
        {Assistant's questions},
        {Read the generated plan},
        {Ran or tested the program},
        {Did not explicitly think about this decision},
        {Read the generated code},
        {\parbox[c]{4.5cm}{\raggedleft\emph{``Read the context/evidence the}\\\emph{AI assistant gave me''}*}},
        {\parbox[c]{4.5cm}{\raggedleft\emph{``I let Claude think of the criteria all together''}*}},
        {\parbox[c]{4.5cm}{\raggedleft\emph{``I asked Claude to come up with}\\\emph{a ranking strategy''}*}}
    },
    yticklabel style={font=\scriptsize, anchor=east},
    y dir=reverse,
    enlarge y limits={abs=0.25cm},
    legend style={draw=none, fill=none},
    legend columns=0,
    axis lines=left,
    axis line style={draw=none},
    ymajorgrids=true,
    grid style={gray!30, thin},
    xtick=\empty,
    tick style={draw=none},
    ytick style={draw=none},
    clip=false,
]

\addplot[fill=aporia-purple, draw=none,
    nodes near coords,
    nodes near coords style={font=\tiny, color=white, anchor=center},
    point meta=explicit symbolic,
] coordinates {
    (15,0) [15]
    (32,1) [32]
    (4,2) [4]
    (5,3) [5]
    (2,4) [2]
    (1,5) [1]
    (1,6) [1]
    (0,7) []
    (0,8) []
};

\addplot[fill=claude-orange, draw=none,
    nodes near coords,
    nodes near coords style={font=\tiny, color=white, anchor=center},
    point meta=explicit symbolic,
] coordinates {
    (26,0) [26]
    (6,1) [6]
    (11,2) [11]
    (9,3) [9]
    (1,4) [1]
    (0,5) []
    (0,6) []
    (1,7) [1]
    (1,8) [1]
};

\end{axis}
\end{tikzpicture}

\captionof{figure}{Participants' self-reported decision-making strategies as described in~\autoref{sec:mental-review}.}
\label{fig:decision-strategies}
\Description{Horizontal stacked bar chart comparing participants using Aporia and Claude Code. Aporia participants most often cited assistant questions as the strategy that helped them notice a design decision, while Claude Code participants most often cited prior knowledge.}
\end{minipage}%
\hfill
{\color{gray}\vrule}
\hfill
\begin{minipage}[t]{0.48\textwidth}
\vspace{0pt}
\begin{flushleft}
\footnotesize
\faClipboardList\;\textbf{Survey Question:} What strategies did you use to validate your understanding of your implementation? (select up to two)

\vspace{2pt}
\end{flushleft}
\scriptsize
\tikz\fill[aporia-purple] (0,0) rectangle (0.2cm,0.2cm);\,\sysplain\quad \tikz\fill[claude-orange] (0,0) rectangle (0.2cm,0.2cm);\,\claudeplain\quad *User-written response

\vspace{4pt}
\hspace{-1.9cm}\begin{tikzpicture}
\begin{axis}[
    xbar stacked,
    width=0.7\columnwidth,
    height=4cm,
    bar width=8pt,
    xmin=0,
    xmax=26,
    ytick=data,
    yticklabels={
        {Manually tested the web app},
        {Read the generated plan},
        {Read the generated code},
        {Ran unit tests},
        {Read the generated tests},
        {Did not verify (trusted the agent)},
        {\parbox[c]{4.5cm}{\raggedleft\emph{``Asked the agent to sanity check}\\\emph{my understanding of the tests''}*}}
    },
    yticklabel style={font=\scriptsize, anchor=east},
    y dir=reverse,
    enlarge y limits={abs=0.25cm},
    legend style={draw=none, fill=none},
    legend columns=0,
    axis lines=left,
    axis line style={draw=none},
    ymajorgrids=true,
    grid style={gray!30, thin},
    xtick=\empty,
    tick style={draw=none},
    ytick style={draw=none},
    clip=false,
]

\addplot[fill=aporia-purple, draw=none,
    nodes near coords,
    nodes near coords style={font=\tiny, color=white, anchor=center},
    point meta=explicit symbolic,
] coordinates {
    (12,0) [12]
    (2,1) [2]
    (2,2) [2]
    (4,3) [4]
    (2,4) [2]
    (1,5) [1]
    (0,6) []
};

\addplot[fill=claude-orange, draw=none,
    nodes near coords,
    nodes near coords style={font=\tiny, color=white, anchor=center},
    point meta=explicit symbolic,
] coordinates {
    (12,0) [12]
    (9,1) [9]
    (4,2) [4]
    (2,3) [2]
    (0,4) []
    (0,5) []
    (1,6) [1]
};

\end{axis}
\end{tikzpicture}

\vspace{1cm}
\captionof{figure}{Participants' self-reported validation strategies. Each bar indicates the number of responses for each strategy.}
\label{fig:validation-strategies}
\Description{Horizontal stacked bar chart comparing participants using Aporia and Claude Code. Both groups most often manually tested the web app; Aporia participants more often ran unit tests, while Claude Code participants more often read the generated plan and code.}
\end{minipage}
\end{figure*}

Nine participants (P1, P4, P7, P8, P9, P10, P11, P12, P16) noted that \sys surfaced considerations they had not thought of.
P4 reflected that the questions \emph{``made me realize {\ldots} edge cases or {\ldots} issues that a user could face while using the program,''} and P1 remarked that without the questions, they would \emph{``miss at least 5 of [their decisions].''}
This observation is supported by participants' self-reported decision-making strategies, shown in~\autoref{fig:decision-strategies}.
When asked ``What helped you realize there was a decision to make here?'' in the
post-task survey, \sys users most frequently reported that the {assistant's questions} brought decisions to their attention,
while \claude users relied mainly on their own {prior knowledge or experience}.

\subsubsection{\sysplain helped participants articulate design decisions from vague intent, but success depended on question quality}
\label{sec:articulate}

Even when participants were already aware of a design consideration,
\sys's questions made it easier for them to articulate and document their decisions.
This is consistent with the principle of recognition over recall~\cite{neilsen1994}:
recognizing a question that aligns with your intent
is easier than recalling and verbalizing that intent. 

We observed \sys surfacing questions that resonated with the existing intuitions of six participants (P1, P4, P11, P12, P16, P19).
For example, while answering questions, P19 remarked \q{I had a specific question in mind.} 
After reading a newly generated question, they followed with \q{Oh \ldots okay, so that question is already generated.}
Reflecting on their task, P12 noted that \emph{``because [\sysinquote] kept lining up with things that I thought were appropriate, {\ldots} I trust this process. It seems to be doing what I would have done anyway.''}

As noted in \autoref{sec:design-considerations},
we chose to make \sys's design question binary as opposed to open-ended.
Five participants (P3, P7, P12, P16, P19) noted that this made it easier for them to quickly approve of decisions that were in line with their thinking.
When asked if they found \sys's questions useful,
P12 said that they appreciated \q{just being able to {\ldots} say yes or no to something, as opposed to coming up with a question and then answering it.}

\sys's questions were not always on the mark.
P7, P8, P10, and P11 found some questions irrelevant, with P7 noting at
different moments that the questions were either ``too detailed'' or ``too obvious.''
When \sys's questions resonated, however,
they seemed to lighten the cognitive demand of planning.
While the NASA-TLX scores did not reveal a significant difference in cognitive load between assistants,
four participants (P3, P4, P9, P12) noted that they felt \sys was less demanding to use than \claude.
P4 said that \sys required less effort because \emph{``it asked me questions about the decisions, and I didn't really have to think about {\ldots} are there decisions to make, and what are they?''}.

In contrast, seven participants (P4, P7, P8, P10, P15, P16, P19) struggled to articulate their decisions when working with \claude.
P8 reflected: \emph{``It's kind of tricky, because [\sysinquote] broke it down into the design decisions for me, but for [\claudeinquote], it's more like everything together.''}

While \claude occasionally asked questions of participants, it did not do so as frequently or systematically as \sys:
we found that on average, participants using \sys made 13.5 times more elicited decisions (95\% CI $[10.9, 16.7]$, $p <0.001$), decisions made in response to an agent question, than those using \claude.
More interestingly, \sys users also made 2.99 times more decisions overall (95\% CI $[2.69,3.32]$, $p<0.001$) than \claude users.
This suggests that elicited decisions are less costly to make than proactive ones,
and that decision elicitation increases overall engagement with the design process.
(We note that there is some subjectivity in how we counted non-elicited decisions with \claude; see \autoref{sec:decision-coding}.)

\subsection{RQ2: Perceived vs. actual understanding}
\label{sec:rq2}

\subsubsection{Participants had a significantly more accurate mental model of their code with \sysplain}
\label{sec:accuracy}
Through our correctness analysis described in~\autoref{sec:correctness} and illustrated in~\autoref{fig:sample-correctness}, we compared participants' descriptions of their implementation (the Description grid) against the code they submitted
 (the Implementation grid).
Participants described a similar number of policies (red or green cells in~\autoref{fig:sample-correctness}) with both assistants ($M_{\sysplain}=10.93$, $M_{\tname{Claude}}=10.43$; $p=0.68$), but their Description grids were 
more accurate with \sys.

For instance, P12 wrote that \emph{``unrelated users can see nothing,''} for
Task A, but their code lacked an access control check on the PDF endpoint,
allowing unrelated users to bypass the policy---this was a {\setlength{\fboxsep}{1pt}\colorbox{cellred}{mismatch}} at the \scode{PDF/Unrelated} intersection in \autoref{fig:sample-correctness}. 
Mismatches like these were 79\% less likely with \sys ($RR=0.21$, 95\% CI $[0.08, 0.55]$, $p<0.001$). 
On average, 4.7\% of described policies were mismatches with \sys vs. 12.1\% with \claude.

Conversely, P3 wrote that \emph{``reviewers from the same university as the authors can't be assigned to their papers''} and their code enforced this ({\setlength{\fboxsep}{1pt}\colorbox{cellgreen}{match}} at the \scode{Different institution} cell in \autoref{fig:sample-correctness}). 
Such matches were 14\% more likely with \sys ($RR=1.14$, 95\% CI $[1.06, 1.23]$, $p < 0.001$). 
On average, 94\% of described policies were successfully implemented with \sys vs. 88\% with \claude.

Finally, P3 did not mention a reviewer ranking strategy, but their code sorted them alphabetically ({\setlength{\fboxsep}{1pt}\colorbox{cellgray}{not described}} at the \scode{Alphabetical} cell in \autoref{fig:sample-correctness}). 
Such undescribed policies were less common with \sys, though this difference was not statistically significant ($M_{\sysplain}=1.71$, $M_{\tname{Claude}}=2.57$; $IRR=0.67$, 95\% CI $[0.40, 1.12]$, $p=0.124$).

\subsubsection{Participants had a more thorough mental model of their design with \sysplain}
\label{sec:thorough}

As described in~\autoref{sec:study}, participants rated their confidence twice: once before and once after completing the systematic mental review described in ~\autoref{sec:mental-review}.
We used the relationship between these two ratings (measured using Spearman rank-order correlation~\cite{spearman}) to assess whether reflection during the post-task surveys changed participants' subjective assessment of their design.
With \sys, we found a significant positive correlation between participants' initial and final confidence ($\rho=0.70$, $p=0.005$), indicating that their confidence did not shift much during the systematic mental review.
We attribute this stability to \sys's emphasis on reflection
  during the whole implementation process; participants had already
thought through many design decisions before they were explicitly prompted to.  

Qualitatively, participants (P1, P3, P4, P7, P9, P10, P16, P17) valued the Decision Bank as a key reflection point, serving as what P9 called \q{a trace of those decisions that were made}.
We observed participants actively returning to the bank to reinforce their understanding of their designs.
For example, P8 walked through each decision while waiting for \sys's
implementation to finish, rephrasing them in their own words: \emph{``Oh, so
  this is for admin, which we weren't really clear on the policies... okay, this
  is kind of covering the thing that a person can have multiple roles... And
  this is for the PDF files? So, I think this is handling the ambiguity of what
  paper details covers.''} 

Participants (P1, P4, P12, P16, P19) also remarked that the breadth of \sys's questions led to more calibrated confidence in their implementation, with P4 noting \emph{``because [\sysinquote] asked me all those questions and I made that decision, I would be pretty confident.''} 
P1 described how the volume of questions facilitated a \emph{``pretty well-rounded implementation''}: \emph{``I would definitely stop after 5 or 6, but it kept generating, so I'm like, oh, maybe this is a good thing to think about too.''}

For participants working with \claude, the correlation between their initial and final confidence was not significant ($\rho=0.26$, $p=0.37$), indicating variability in their confidence upon reflection. 
For some participants, the post-task survey was the first time they reflected on their design at all, and seven (P3, P7, P8, P10, P16, P18, P19) verbalized doubts when having to describe their design (see~\autoref{fig:sample-correctness}).
P3 noted mid-survey: \emph{``I'm thinking oh, the plan was not as good... because there's so many statuses, like, before submission, after submission, and, for each role...I was not careful enough to approve the plan.''}
P8 echoed, \emph{``There's some decisions I feel like I didn't really
    get into... I just [thought] of that when I was filling out the form.''}
P10 simply said, \emph{``I think I should have not, uh, I think I should have spent more time.''} when questioning whether reviewers can see other reviews. 
This phenomenon was not observed with \sys.

\subsection{Validating the design motivations}
A mixture of study results and participant feedback offer evidence of the role that our design motivations for decision-oriented programming played in the usefulness of \sys:

\subsubsection{Explicit and structured decisions} \label{sec:DG1}
DOP's goal of reifying design decisions (\hyperref[dg:DG1]{\textbf{DG1}}) is operationalized by \sys's Decision Bank, which was the most positively received of \sys's affordances (\autoref{fig:likert-aporia}); ten participants (P1, P3, P4, P7, P8, P9, P10, P11, P16, P17) specifically credited it with helping them keep track of what they had decided.

P17 appreciated it as \emph{``a \textbf{quick mental map} of the overarching decisions that we've made.''}, and P7 valued being able to \q{just go back to it at any time}.
Five participants (P1, P3, P16, P17, P19) found \claude's unstructured chat history harder to understand than \sys's Decision Bank.
P16 noted that \sys \q{definitely felt more systematic{\ldots}I felt more confident about what [\sysinquote] was doing{\ldots}reading through a plan on text is a lot of effort.}
P19 echoed this point, adding \q{I wish I had something like this to use in my work.}

Participants took advantage of decisions being laid out in front of them in a variety of ways.
P3 came back to the Bank to confirm that their decisions were not in conflict.
P11 used the Bank to step back and review their plan after making a few decisions in quick succession.
Several participants (P3, P4, P10, P16) appreciated the ability to edit decisions in the Bank, and three (P3, P8, P17) did so during the study.

In contrast, five participants (P4, P7, P8, P9, P16) struggled to recall key aspects of their \claude chats.
P9 described \q{so much{\ldots} chain of thought being generated{\ldots}you just can't really always see previous decisions,}
and P3 noted that with everything mixed together in the chats, \q{it's not easy for me to go back to find the information I need. {\ldots} [\sysinquote] collects all the same type of information together, so it's better.}
Prior studies corroborate this difficulty, finding that developers resort to organizing their thoughts in separate files to compensate for unstructured chat interfaces~\cite{huang-2025}.

\subsubsection{Interactively co-authored decisions} \label{sec:DG2}
DOP's goal of encouraging interactive decision-making (\hyperref[dg:DG2]{\textbf{DG2}}) is operationalized by \sys's Question Bubbles, which generally led participants to think more critically during feature design, but did not guarantee engagement.
P4 noted that \q{[\sysinquote] makes you think about things that the AI would have normally just made its own assumptions \ldots [you never] would have had an opportunity to \ldots make those decisions yourself.}
P1 contrasted this sentiment, \q{[\claudeinquote] removes a lot of thinking, \ldots I don't think anymore\ldots I just prompt}.

Five participants (P1, P4, P9, P11, P12) said that \sys's questions made them think more deeply about decisions they had been considering.
P9 reflected on a decision they made about reviewer ranking,
\emph{``I was bringing in my own experience...the question the assistant asked me was{\ldots} asking me to think deeper myself.''}

When participants were unsure how to answer a question,
\sys's Question Detail View (\autoref{fig:user-scenario}, \circledletter{3}) provided the necessary context with rationale and code references.
After seeing the Detail View, P3 noted that normally, when using \claude, they \emph{``explicitly asked [the agent] to think about the rationale before the decision. But here{\ldots} I don't need to explicitly ask.''}
Five participants (P4, P7, P9, P12, P19) used the relevant code references to ground their decisions in the codebase.
In one case, P4 was directly influenced by seeing \sys's reference to existing code that instantiated a three-reviewer limit on papers: \emph{``normally{\ldots} I would have said no{\ldots} but I guess since the code already has these constraints that it told me about{\ldots} I'm gonna go with what it says.''}

Together, questions and grounding helped participants reflect on their decisions in the context of the codebase.
For example, P11 had never used a paper management system and initially assumed that \q{There is, like a class of{\ldots} admins{\ldots} who are{\ldots} the reviewers of all the papers},
but upon reflection, decided that: \q{Basically, anyone can probably be a reviewer, anyone can be a submitter, and then that's completely separate from who's an admin.}

We did observe cases where participants did not think deeply about their decisions or ground them in the codebase using \sys.
P9 demonstrated two failure modes: they made a custom decision to \q{Exclude
  people who provided mentorship and/or funding to the paper authors} despite no
data about mentorship or funding existing in the codebase (no grounding) and later admitted \q{I was just kind of clicking yes on the
  assistant for some of them, {\ldots} I wasn't thinking as deeply}.
These instances illustrate that the Question Bubbles did not guarantee reflection in all cases.

\subsubsection{Traceability between decisions and code} \label{sec:DG3}
DOP's goal of tracing decisions to code (\hyperref[dg:DG3]{\textbf{DG3}}) is operationalized by \sys's test suites, which helped a subset of participants validate that the agent's implementation reflected their intent.
However, test suite reception was mixed and many participants lightly engaged with tests regardless of assistant.

As shown in \autoref{fig:validation-strategies}, participants using both
assistants reported manual testing as their primary strategy for
checking that their \notcrp code aligned with their intent.
Their secondary strategies differed, though: \sys participants tended to rely on test suites, while \claude participants relied on reading the generated plan and code.

Additionally, although participants rarely explicitly read the test suites in the way we expected, we observed several informal uses of the tests that helped validation.
For example, \sys's test suites often directly influenced how participants approached manual testing.
Five participants (P4, P8, P12, P15, P17) mapped test suites back to the Decision Bank to validate that their intent was reflected.
P4, while reviewing the generated tests, noted \emph{``Okay, yeah, these seem like they're related to the answers I selected.''} 
P15 similarly observed that \emph{``the comments are very clear about what's
  happening... I can kind of map it back to the questions that I answered over here.''}

Five participants (P3, P4, P8, P12, P16) also used \sys-generated test suites to \emph{identify what to validate}.
For example, before testing new functionality in \notcrp, P12 
said \emph{``What did I add since last time?... [reads test
  suite names aloud]... Okay, so I should go to the draft [page].''}
Afterwards, they reflected that the test suites \emph{``complement...
  looking at the website... if I want to check a
  specific decision... it's nice to be able to \textbf{pinpoint where this came
    from and why it exists}... and instead of looking at test names and kind of
  guessing what it is, it says...I did this for this reason.''}

Despite these use cases, a majority of participants (P1, P3, P4, P7, P8, P9,
P12, P15, P16, P17, P18) stated that they did not always engage with the test
suites produced by either assistant.  
P17 put it most directly: it was
\emph{``very useful that \sys generated [test suites]. But I didn't really use
  them.''}  
P7 found the tests too verbose: ``\emph{it's not so helpful [because I don't] inspect [the tests] in\ldots detail, and it's hard to...verify...whether it is actually correct.} 
Instead of using the test suites to aid their thought process, many saw the tests more as a \emph{``mechanized way for the agent to double-check itself''} (P9).

Participants using \claude relied primarily on reading the generated plan and code to supplement manual testing, but as discussed in~\autoref{sec:DG1}, many participants expressed difficulty in locating and tracking their decisions within the plan. 
\claude only generated test suites for five participants (P4, P8, P12, P15, P18), and all but one had to explicitly prompt the agent to do so. 
As with \sys, participants did not engage with the tests in detail---\emph{``I
  told [\claudeinquote] to make tests, and then I didn't really look at the tests''}
(P15)---but without traceability to the plan, the tests offered little
scaffolding or support for validation.
\subsection{Limitations}
Our study has two main threats to external validity,
the first being the targeted scope of our tasks. 
We deliberately chose open-ended, decision-dense tasks to isolate DOP's effect on programmer decision-making.
We accordingly do not claim that DOP is well-suited for smaller, more well-defined tasks (i.e. bug fixing or refactoring).
The second limitation is the limited time spent on each task, which constrained the scope of what participants could implement.
A longitudinal study could further investigate how DOP can benefit programmers as a codebase and their goals evolve.
\highlighttoolsfalse
\section{Discussion \& Future Work}

\subsection{When is decision-oriented programming worth it?}
Our results show that DOP's central affordance of eliciting and reifying decisions encourages programmers to slow down, articulate their intent more precisely, and inspect decisions they might otherwise defer to an agent, but this effort is not universally desirable.
Developers often use coding agents to reduce articulation effort and stay in flow~\cite{fawzy-vibe-coding}.

In some real-world vibe coding scenarios, users intentionally avoid specifying everything upfront when in an exploratory phase of development. 
Upon reflection, some participants said they preferred this workflow, or a mix of both upfront intent specification and support for iteration.
Many participants said that they would prefer DOP when ``uncertain about what they wanted'', or for very critical tasks, and a more traditional coding agent when they had ``clear goals'' or for ``small edits'' that they wanted to finish quickly. 
This suggests that DOP is most beneficial when program requirements or developer intent are unclear: precisely when decisions are ordinarily delegated silently to agents.

The success of DOP-aligned systems also depends on the way they represent decisions.
Our probe treats decisions as atomic, behavioral, and testable choices that clarify a user's intent for a feature (\autoref{sec:design-considerations}). This is only one approach to representing decisions; they could instead be relational, non-behavioral, or validated by other methods.
Because each decision is encoded as a test, our probe can only capture decisions with testable behavior.
Representing decisions as atomic and testable fits the long tail of software development, like incremental additions to web applications from our study, more than domains like optimization algorithms or compilers where behavior is tightly interdependent.

\subsection{Decision elicitation risks and rewards}
Study observations suggest that the success of DOP's decision elicitation mechanism depends on the quality of the questioner (the agent, in our case).
Our qualitative results identify \emph{productive} interruptions, where the agent's questions surfaced helpful ideas participants hadn't considered or aligned with their current thinking, and \emph{unproductive} ones, where questions were overly detailed or irrelevant (\autoref{sec:articulate}).

A subtler risk of decision elicitation is that the questioner may fail to ask valuable questions.
Identifying critical code generation decisions (and ensuring they are elicited) is an open problem also posed in recent work~\cite{hilde}. 
We attempted to mitigate this risk by letting users add their own decisions and modify agent generated decisions, but helping programmers identify decisions the agent has not surfaced is a fruitful avenue for future work.

Another risk is that the questioner may elicit decisions based on a hallucinated state of the codebase, though we did not observe this in our study.
Our probe added relevant code snippets to each question to help ground decisions in context, but a more systematic approach could draw on program analysis or formal methods to guarantee that a question refers to a state that exists in the code. 
Recent work moves in this direction, symbolically deriving multiple choice questions and scenarios that disambiguate user intent and using an LLM only to phrase them in natural language~\cite{barnaby2026choose, prasad2026meaningful}. 

Finally, binary questions are only one way to elicit decisions, and our probe did not always strike the right balance between friction and flow (see ~\autoref{sec:DG2}), as some participants tended to answer ``yes'' rather than engaging critically with a decision.
We are interested in exploring other UI techniques such as selecting from a list of alternatives or inferring intent from manual code edits and chats, inspired by Programming by Demonstration~\cite{albarghouthi2013recursive, gulwani2011automating}.

\subsection{The importance of skimmable representations of intent}
Our design probe gave study participants a two-sided representation of their intent: an informal list of design decisions (the Decision Bank) and the formal test suites that validate those decisions in the context of the codebase (see~\autoref{sec:aporia}).
We had expected participants to make use of the test suites, and were surprised to see that they rarely engaged with them (see~\autoref{sec:DG3}). 
Participant comments suggest that the tests were too verbose to read closely, in line with findings that developers have difficulty scrutinizing LLM-generated tests~\cite{kaufman2026programmers}.
We tried to offset this difficulty by adding natural-language comments summarizing each unit test's inputs and outputs, but they were frequently overlooked during the tasks.
Instead, participants repeatedly relied on the Decision Bank (see~\autoref{sec:DG1}); as one commented, compared to a natural-language plan it \emph{``is better for visual search... it has more colors, and things are organized in chunks.''}

We believe what set the two representations apart was \emph{skimmability}.
As we explore representations of user intent beyond flat lists of decisions or test suites, we (and future system designers) should keep in mind that developers will not use these affordances in a fast-paced agentic environment unless they are understandable at a glance.
This reflects the cognitive dimension of \emph{synopsie}, a notation's ability to convey the whole when you ``stand back and look''~\cite{blackwell2003notational}.

\subsection{Supporting structure without premature commitment}
Our probe scaffolds decision-making around an explicit goal and a collection of user decisions underneath before generating any code, which gave our participants more opportunities to reflect during implementation (see~\autoref{sec:DG2}).
However, because our probe was scoped to a single design session, it offered incomplete support for revising a goal mid-task.
Three participants had difficulty ``re-steering'' the probe as a result, and two reported that they preferred the ``streamlined flow'' of a conventional agent, consistent with findings that programmers following predefined strategies see their work as more organized but more constrained~\cite{LaTozaALK20}.
This rigidity can be classified as what the cognitive-dimensions framework calls \emph{premature commitment}, where a notation constrains the order in which one can act~\cite{blackwell2003notational}.

Future DOP systems could reduce premature commitment by helping programmers \emph{reassess} decisions after making them. 
During our study, for instance, the probe occasionally found flaws in its own tests but left them unedited to preserve user intent; 
instead, it could surface such a flaw as a new question, letting the programmer iterate on how their intent is interpreted by the agent.
Additionally, we built our probe to study the DOP workflow in isolation, not to compete with conventional coding agents, so it lacks free-form conversational interaction. 
Future systems could explore how to combine the structure of DOP with the flexibility of conventional agents, so that programmers can move fluidly between structured decision-making and open-ended chat.


\subsection{If decisions are objects, we can reason over them!}
Our study showed that atomic decisions collected linearly in a Decision Bank gave participants a useful structure for solidifying their intent (\autoref{sec:DG1}), and in future iterations of DOP, we are interested in other approaches to organizing and reasoning over user decisions.

We observed that participants were sometimes unsure of what order to answer questions in, how decisions related to one another, or when they had made ``enough'' decisions to fully specify their goal and begin implementation.
These difficulties suggest that DOP could provide more structure to the decision space by organizing decisions hierarchically through follow-up questions, or as a graph that highlights the dependencies and causal relations between decisions.
For example, if a user decides that only a paper's authors and assigned reviewers can view a paper's detail page, the system can follow up with questions about how to handle unauthenticated requests and whether the same policy should apply to other endpoints where a paper's data is exposed.

Other participants wondered about a ``saturation point'' where most key decisions around a feature had been addressed, motivating a UI indicator that suggests related features to scope out next. 
Similarly, participants suggested a navigable ``goal history'' that clusters decisions under different higher-level goals, giving programmers an editable record of how their intent evolved while implementing a larger feature.
We are inspired by recent work that detects semantic conflicts between a user's decisions when planning with AI~\cite{semantic-commit}. Such conflict detection is also possible in software design once user decisions are made into first-class objects rather than being left implicit in a chat history.
\section{Conclusion}
In this work, we explored decision-oriented programming (DOP), a workflow which seeks to support programmer decision-making in agentic software design by reifying decisions as first-class objects.
We instantiated DOP in a design probe called \sys, which elicits design decisions from the programmer with questions, tracks them, and encodes them as test suites.

After evaluating \sys in a user study, we found that
	tracking decisions helped participants externalize their mental model of the program design space;
	asking questions helped participants discover and articulate design decisions; and
	encoding decisions as test suites scaffolded participants' validation process.
Through a shared record of decisions, participants stayed engaged in the design process and kept their understanding aligned with what the agent implemented.

Our experience building and evaluating \sys indicates that DOP's design goals have the potential to support programmer decision-making and promote user agency while retaining many of agentic programming's advantages.
We are excited to continue exploring how best to instantiate decision-oriented programming in future research.

\begin{acks}
This work was supported in part by the NSF under Grant No. CCF-2107397 and by the Amazon Research Award program. 
This material is based upon work supported by the National Science Foundation Graduate Research Fellowship under Grant No. DGE-2038238. 
Any opinions, findings, and conclusions or recommendations expressed in this publication are those of the authors, and do not necessarily
reflect the views of the sponsoring entities.

We would like to thank Carlo Furia, Brian Hempel, Aaron Broukhim, Devamardeep Hayatpur, and Matthew Beaudouin-Lafon for their invaluable feedback.
\end{acks}

\bibliographystyle{ACM-Reference-Format}
\bibliography{refs}


\end{document}